\documentclass[aps,prl,reprint,showpacs,superscriptaddress,amsmath]{revtex4-1}
\usepackage{color}
\usepackage{graphicx}
\begin{document}

\title{Origin of the Resistivity Anisotropy in the Nematic Phase of FeSe}

\author{M.~A.~Tanatar}
\email[corresponding author: ]{tanatar@ameslab.gov}
\affiliation{Ames Laboratory, Ames, Iowa 50011, USA}

\affiliation{Department of Physics and Astronomy, Iowa State University, Ames, Iowa 50011, USA }

\author{A.~E.~B\"ohmer }
\affiliation{Ames Laboratory, Ames, Iowa 50011, USA}

\author{E.~I.~Timmons}
\affiliation{Ames Laboratory, Ames, Iowa 50011, USA}
\affiliation{Department of Physics and Astronomy, Iowa State University, Ames, Iowa 50011, USA }

\author{M.~Sch\"utt}
\affiliation{School of Physics and Astronomy, University of Minnesota, Minneapolis, Minnesota 55455, USA}

\author{G.~Drachuck}
\affiliation{Ames Laboratory, Ames, Iowa 50011, USA}
\affiliation{Department of Physics and Astronomy, Iowa State University, Ames, Iowa 50011, USA }

\author{V.~Taufour }
\affiliation{Ames Laboratory, Ames, Iowa 50011, USA}

\author{S.~L.~Bud'ko }
\affiliation{Ames Laboratory, Ames, Iowa 50011, USA}
\affiliation{Department of Physics and Astronomy, Iowa State University, Ames, Iowa 50011, USA }

\author{P.~C.~Canfield}
\affiliation{Ames Laboratory, Ames, Iowa 50011, USA}
\affiliation{Department of Physics and Astronomy, Iowa State University, Ames, Iowa 50011, USA }

\author{R.~M.~Fernandes}
\affiliation{School of Physics and Astronomy, University of Minnesota, Minneapolis,
Minnesota 55455, USA}

\author{R.~Prozorov}
\email[corresponding author: ]{prozorov@ameslab.gov}
\affiliation{Ames Laboratory, Ames, Iowa 50011, USA}
\affiliation{Department of Physics and Astronomy, Iowa State University, Ames, Iowa 50011, USA }

\date{15 November 2015}

\begin{abstract}
The in-plane resistivity anisotropy is studied in strain-detwinned single crystals of FeSe. In contrast to other iron-based superconductors, FeSe does not develop long-range magnetic order below the nematic/structural transition at  $T_{s}\approx$90~K. This allows for the disentanglement of the contributions
to the resistivity anisotropy due to nematic and magnetic orders. Comparing direct transport and elastoresistivity measurements, we extract the intrinsic resistivity anisotropy of strain-free samples.  The anisotropy peaks slightly below $T_{s}$ and decreases to nearly zero on cooling down to the superconducting transition. This behavior is consistent with a scenario in which the in-plane resistivity anisotropy in FeSe is dominated by inelastic scattering by anisotropic spin fluctuations.
\end{abstract}


\maketitle

Electronic nematicity has emerged as a key concept in iron-based superconductors
since the observation of in-plane resistivity anisotropy in
stress-detwinned crystals of Co-doped BaFe$_{2}$As$_{2}$ \cite{detwinning,Fisher1}.
The fact that the resistivity anisotropy is much larger than what is expected from the small lattice distortion led to the proposal that the tetragonal-to-orthorhombic transition in the iron pnictides is driven not by phonons, but by an electronic nematic phase. Subsequent experiments revealed an intricate dependence of the resistivity anisotropy on doping (a sign change between electron- and hole- doped materials \cite{Fisher1,Chen,ErickNature,Chen2,Miyasaka}), and disorder \cite{Uchida,Fisher14}, sparking hot debates about its microscopic origins (see Refs.~[\onlinecite{FisherReview,FernandesNaturereview}]
for reviews).

Electronic contributions involved in the in-plane resistivity
anisotropy \cite{FernandesNaturereview} can be separated into the Drude weight and/or of the scattering rate anisotropies. Fermi-surface anisotropies arising, for
instance, from the ferro-orbital order triggered at the nematic transition,
affect mostly the Drude weight \cite{Devereaux10,Phillips11,Dagotto12}.
Anisotropic scattering, can be due to elastic processes, such as the development of local magnetic order around
an impurity \cite{Davis13,Andersen}, or inelastic processes,
such as the scattering of electrons by anisotropic magnetic fluctuations
\cite{Rafaelcalc,Timm14} known to exist below $T_{s}$ \cite{Dai14}.
Recent stress-dependent optical reflectivity studies in
Co-doped BaFe$_{2}$As$_{2}$ point to a dominant
effect of the Drude weight \cite{Mirri2014,Mirri2015}.
However, stripe magnetic order
appearing at the magnetic transition severely complicates the analysis. This is because the magnetic state breaks tetragonal symmetry leading to an anisotropic reconstruction
of the Fermi surface \cite{Uchida,Valenzuela10} and to the appearance of
``Dirac cones'' \cite{Dirac}, which may dramatically
alter the resistivity anisotropy \cite{Fisherdiracnematicity}.
Disentangling these contributions is fundamental to reveal the origin of the resistivity anisotropy and, consequently, of the nematic state.

In this context, the stoichiometric FeSe \cite{FeSe} is an ideal system. It is rather clean (residual resistivity ratios as high
as 50 \cite{rrrFeSe}) and its orthorhombic/nematic
phase transition at $T_{s}\approx90$~K is not accompanied by a long-range
magnetic order \cite{McQueen2009} eliminating effects of Fermi
surface folding.

In this Letter we report the resistivity anisotropy measured in
strain-detwinned single crystals of FeSe. Upon cooling, the anisotropy $\Delta\rho(T)\equiv\rho_{a}-\rho_{b}$  ($\rho_a$ and $\rho_b$ are the resistivities along the orthorhombic $a-$ and $b-$ directions) initially increases, reaching a maximum at about 20 K below $T_{s}$, and then nearly vanishes upon further cooling towards the superconducting transition $T_{c}\approx$8.5~K. This pronounced non-monotonic behavior is consistent
with the scenario in which the main contributor
to the resistivity anisotropy
is inelastic scattering by magnetic fluctuations rather than the anisotropy of the elastic scattering or of the Fermi surface. To support this conclusion, we performed model calculations of resistivity anisotropy for the scattering of electrons by anisotropic magnetic fluctuations. We find that the anisotropy is well described by the product of two temperature-dependent functions, $\Delta\rho(T) = \Upsilon(T) \phi(T)$. The standard nematic order parameter, $\phi(T)$, increases upon cooling and the scattering function, $\Upsilon(T)$, decreases and vanishes as the temperature approaches zero.

\begin{figure}
\includegraphics[width=8.2cm]{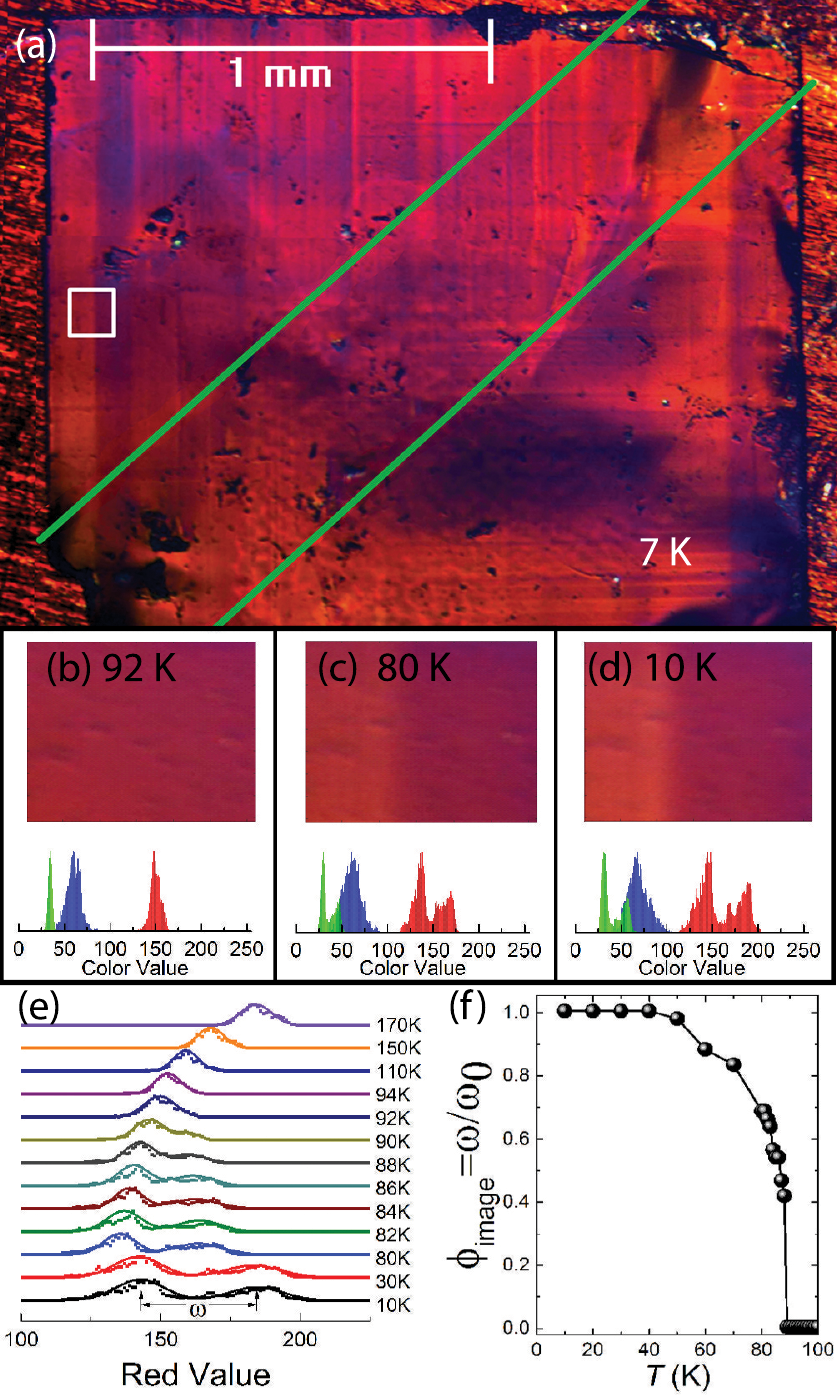} \protect\protect\caption{(color online) (a) Polarized light image of FeSe single crystal at
7 K, revealing orthorhombic domains oriented along tetragonal $[100]$ direction (parallel
to the sample sides). For detwinning, the sample is cut along the
[110] tetragonal direction as indicated by the green lines.
Enlarged are the views of the area indicated by the white box in (a) and corresponding RGB histograms taken at (b) 92 K, (c) 80 K and (d) 10 K. The change below $T_s$ is most pronounced in the red channel (b-d) and temperature evolution of its histograms is shown in panel (e). The peak splitting $\omega$ was analyzed using a fit to two gaussians (lines in (e)) and   normalized nematic order parameter, $\phi_{\mathrm{image}}\equiv\omega(T)/\omega(T \to 0)$, is shown in panel (f).}
\label{imaging}
\end{figure}

Single crystals of FeSe ($\sim$1 mm$^{2}$ surface area and 20 to 150 $\mu$m thick) were grown using a modified chemical vapor transport as in Ref. \onlinecite{Boehmer2013}.
Polarized light optical imaging \cite{imaging,detwinning} was used to characterize
the orthorhombic domain pattern  appearing below
$T_{s}$ as shown in Fig.~\ref{imaging}. In the orthorhombic
phase, the optical bireflectance is anisotropic, which permits visualization of domains of different orientations. In addition to direct imaging, we extract the nematic order parameter from the temperature evolution of the color intensity in a small and clean area of the sample (white box in
Fig.\ref{imaging}(a)), using a decomposition in red, blue and green (RGB) channels \cite{stress}. The intensity histograms of the RGB channels are shown in Figs.\ref{imaging}(b), (c), (d). Above $T_s$ the image is of uniform color, manifesting as single peaks in the histograms (panel (b)). Below $T_s$  the domains of different colors lead to peak splitting in the histograms, most pronounced in the red channel (Figs.\ref{imaging}(c), (d)). The temperature evolution of the red channel histogram is shown in panel (e). The peak splitting, $\omega$, signaling the breaking of tetragonal symmetry, was determined using a fit to two gaussians and the normalized nematic order parameter, defined as $\phi_{\mathrm{image}}(T)\equiv\omega(T)/\omega(T \to 0)$, is shown in Fig.\ref{imaging}(f).

\begin{figure}
\includegraphics[width=8.6cm]{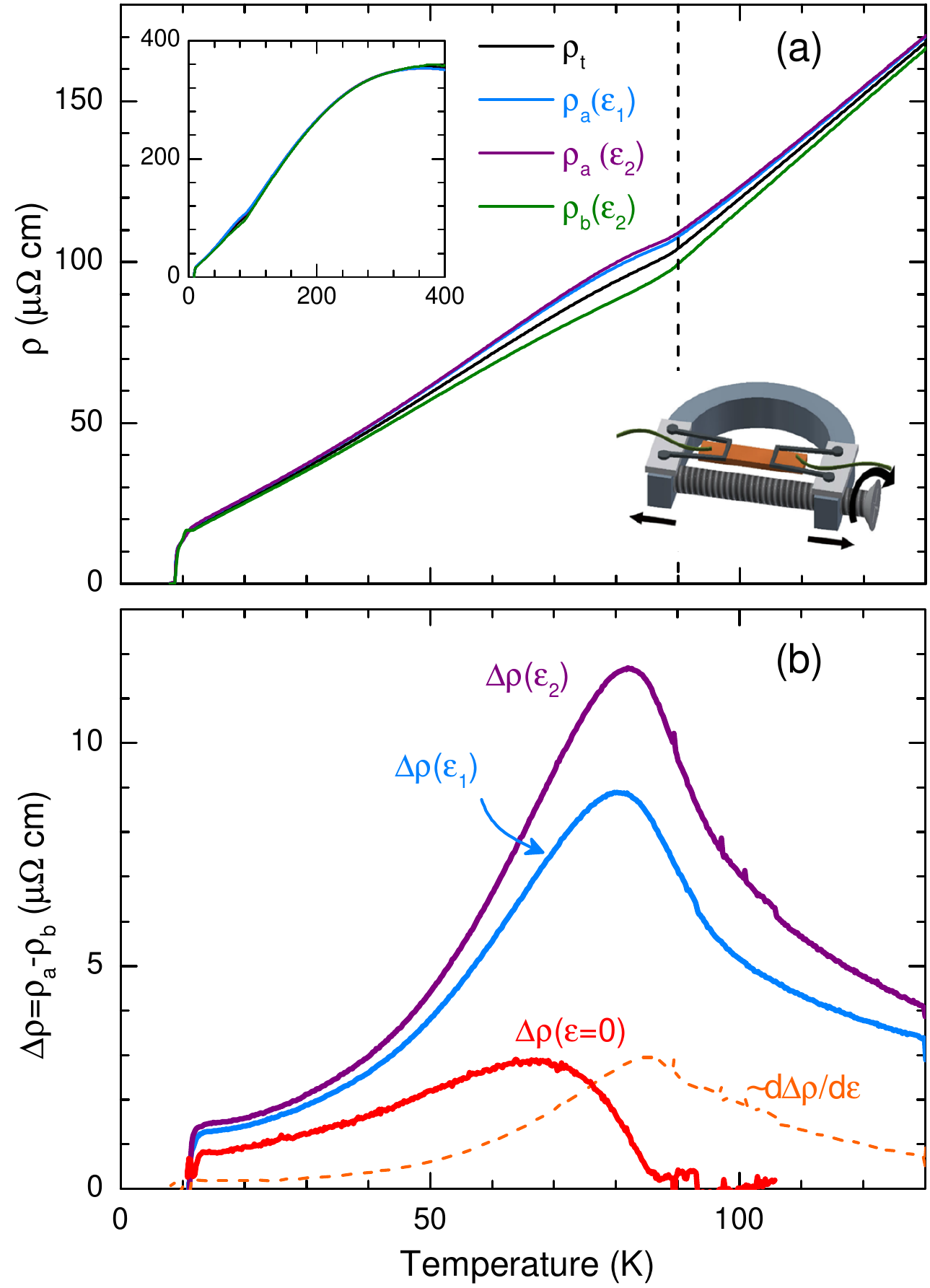} \protect\protect\caption{(color online) (a) Temperature dependent resistivity of FeSe measured in a free-standing state, $\rho_{t}$ (black curve),
and under two values of uniaxial tensile strain, $\varepsilon_a=\varepsilon_{1},\varepsilon_{2}$,
representing a fully detwinned state, $\rho_{a}$ (blue and purple curves). The
resistivity along the orthorhombic $b$ direction was calculated as
$\rho_{b}=2\rho_{t}-\rho_{a}$ (green curve). The insets show the whole temperature range and the schematics of the horseshoe detwinning device. (b) Resistivity
anisotropy, $\Delta\rho\equiv\rho_{a}-\rho_{b}$ for the two values
of strain, $\varepsilon_1$ and $\varepsilon_2$. Their difference $\Delta\rho(\varepsilon_{1})-\Delta\rho(\varepsilon_{2})$ is
proportional to the strain-derivative $d\Delta\rho/d\varepsilon_{a}$
(dashed orange line). The latter was used to extract the intrinsic (strain-free) in-plane
anisotropy of the resistivity $\Delta\rho(\varepsilon_a=0)\approx\Delta\rho(\varepsilon_{2})-\frac{d\Delta\rho}{d\varepsilon_{a}}\,\varepsilon_{2}$ in the
orthorhombic phase (red line).}
\label{resistivity}
\end{figure}

Samples for mechanical detwinning were cut along the tetragonal {[}110{]}
direction, which becomes the orthorhombic $a$ or $b$ axis on cooling,
as shown schematically by the green lines in Fig.~\ref{imaging}(a).
Tensile strain was applied to the sample through 50 $\mu\mathrm{m}$ Ag wires also used as potential leads,
inset in Fig.~2(a). Wires for current contacts
were mounted strain-free. All contacts were In - soldered. Figure~\ref{resistivity}(a) shows the resistivity of the FeSe sample measured in the strain-free
twinned state, $\rho_{t}$, and in the detwinned state achieved by
application of tensile strain $\varepsilon_{a}$ of two different
magnitudes, $\rho_{a}(\varepsilon_{1})$ and $\rho_{a}(\varepsilon_{2})$.
$\varepsilon_{a}=\varepsilon_{1},\varepsilon_{2}$ is controlled by
pulling apart the arms of the horseshoe device. In the strain-free, twinned state $\rho_{t}$ shows
only a small kink at $T_{s}$. The sample is
split into approximately equal areas of domains of two orientations, so its resistivity is $\rho_{t}=(\rho_{a}+\rho_{b})/2$.
Together with the measurements in detwinned samples, this allows us
to extract the resistivity along the orthorhombic $b$ axis, $\rho_{b}$, and the in-plane anisotropy, $\Delta\rho$, shown in
Fig.~\ref{resistivity}(b) for two strain values. The anisotropy increases markedly on cooling, evolves smoothly through $T_{s}$ and peaks below
$T_{s}$. On further cooling it decreases reaching small values at $T_c$.  Note that
$\Delta\rho>0$, i.e., the resistivity is larger along the $a$ direction, thus having the same sign as that
of FeTe \cite{FeTe} and hole-doped BaFe$_2$As$_2$ compounds \cite{ErickNature},
and opposite to electron-doped and isovalently substituted BaFe$_{2}$As$_{2}$ \cite{detwinning,Fisher1,detwinningPdoped}.

\begin{figure}
\includegraphics[width=8.6cm]{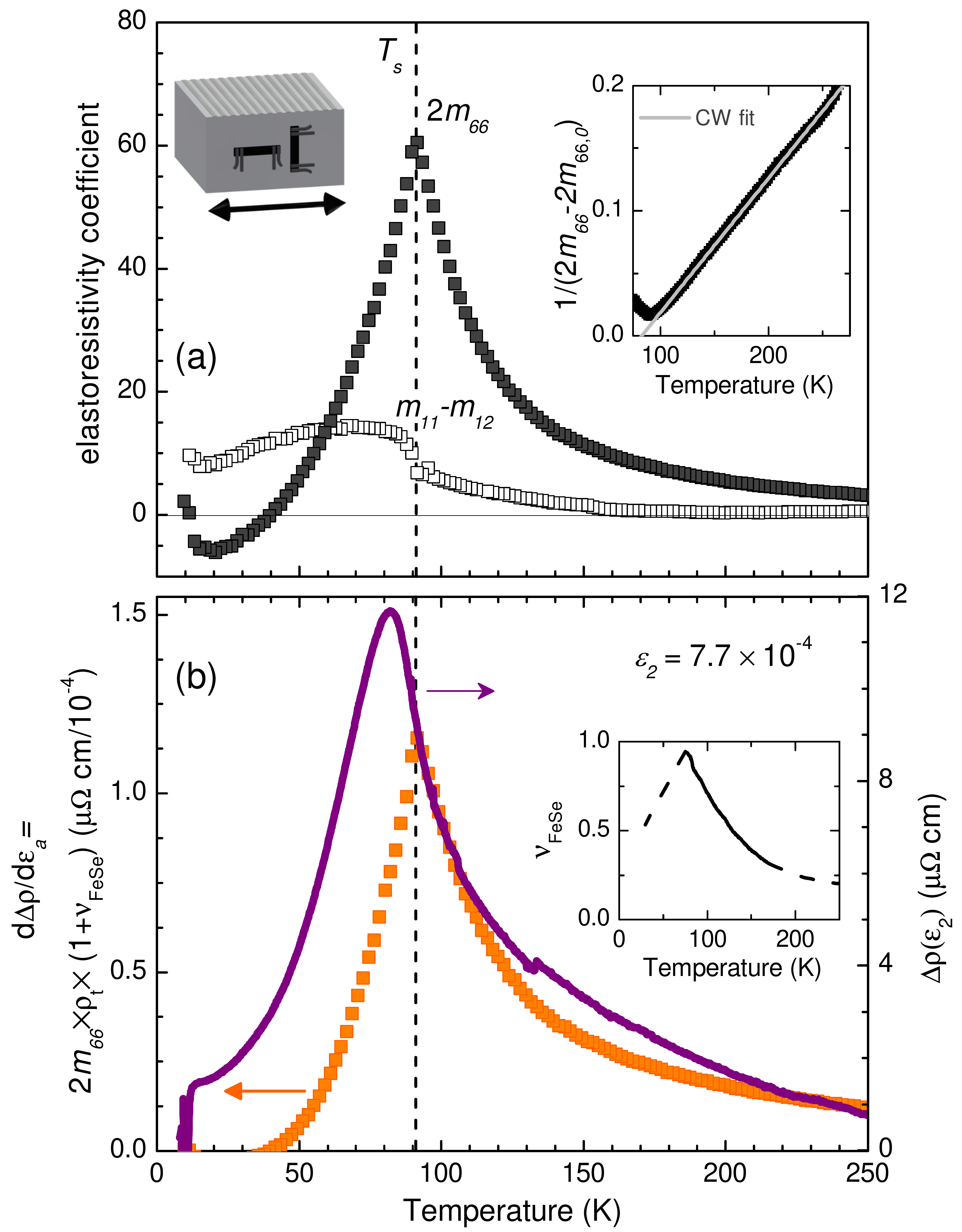} \protect\protect\caption{(color online) (a) Elastoresistivity coefficients $2m_{66}$ and $m_{11}-m_{12}$
of FeSe measured using crossed samples glued to a piezostack, shown schematically in the left inset. The right
inset shows the inverse of $m_{66}-m_{66,0}$ demonstrating nearly
perfect Curie-Weiss-like behavior, $m_{66}=A/(T-T_{0})+m_{66,0}$
with $m_{66,0}=-2.6$ and $T_{0}=83$ K. (b) Scaling of $2m_{66}=\frac{1}{\rho_{t}}\frac{d\Delta\rho}{d\left(\varepsilon_{a}-\varepsilon_{b}\right)}$
with the resistivity anisotropy, $\Delta\rho(\varepsilon_{2})$, measured in detwinned samples. Here, we assume that $\Delta\rho(\varepsilon_{2})$ is induced by the applied strain above $T_s$, so that $\Delta\rho(\varepsilon_{2})=\varepsilon_{2}\left(\frac{d\Delta\rho}{d\varepsilon_a}\right)$ and use the identity $\frac{d\left(\varepsilon_{a}-\varepsilon_{b}\right)}{d\varepsilon_{a}}=1+\nu_{\textnormal{FeSe}}$ to transform between the two quantities. The inset shows the Poisson ratio of FeSe $\nu_{\textnormal{FeSe}}$  determined from the ultrasound data of Ref.~[\onlinecite{Zvyagina2013}] (solid line), and extrapolated to 250 K and below $T_s$ (dashed lines).}

\label{piezo}
\end{figure}

The application of strain not only promotes the formation of orthorhombic domains of only one orientation below
$T_{s}$, but it also induces resistivity anisotropy above and below $T_s$ due to the elastoresistivity of the material. Figure \ref{piezo}(a)
shows the elastoresistivity coefficients 2$m_{66}$ and $m_{11}-m_{12}$
measured using a piezo-based setup, similar to that described
in Refs.~[\onlinecite{Chu2012,nematicsusceptibility}]. Samples of approximate
dimensions, 1$\times$0.3$\times$0.07 mm$^{3}$, were glued to one side
of a piezostack, shown in the left inset in Fig.~\ref{piezo}(a). The change of sample resistance was measured as a function of anisotropic strain, monitored \textit{in
situ} using crossed strain gauges glued to the opposite side of the
piezostack. The elastoresistivity coefficient $2m_{66}$
corresponds to the normalized derivative of $\Delta\rho$ with respect
to the orthorhombic shear strain $\varepsilon_{a}-\varepsilon_{b}$, $2m_{66}=\frac{1}{\rho}\frac{d\Delta\rho}{d\left(\varepsilon_a-\varepsilon_b\right)}$ \cite{nematicsusceptibility}.
It clearly diverges on approaching $T_{s}$ from above (Fig.~\ref{piezo}(a)) following almost perfect Curie-Weiss law, $2m_{66}\sim1/(T-T_{0})$ with $T_{0}\approx$83
K (right inset in Fig.~\ref{piezo}(a)), in qualitative agreement with previous report \cite{Watson2015}.
The elastoresistivity mode, $m_{11}-m_{12}$, ``orthogonal'' to
$2m_{66}$, is related to the derivative of the resistivity anisotropy between two
diagonals of the orthorhombic unit cell, $[110]_\textnormal{o}$ and $[1\bar{1}0]_\textnormal{o}$, with respect to the corresponding shear strain, $m_{11}-m_{12}=\frac{1}{\rho}\frac{d\left(\rho_{[110]_\textnormal{o}}-\rho_{[1\bar{1}0]_\textnormal{o}}\right)}{d\left(\varepsilon_{[110]_\textnormal{o}}-\varepsilon_{[1\bar{1}0]_\textnormal{o}}\right)}$ \cite{nematicsusceptibility}. This mode does not couple to the nematic order parameter and is, as expected, almost zero above $T_{s}$. In the strain-free samples, $\Delta\rho=0$ is expected for $T>T_s$ and the observed finite resistivity anisotropy is likely a consequence of the applied strain. We therefore compare in Fig.~\ref{piezo}(b) the resistivity anisotropy under applied strain $\epsilon_2$, given in this case by $\Delta\rho(\varepsilon_{2})=\varepsilon_{2}\left(\frac{d\Delta\rho}{d\varepsilon_a}\right)$, with the elastoresistivity data. Because $2m_{66}=\frac{1}{\rho_{t}}\frac{d\Delta\rho}{d\left(\varepsilon_{a}-\varepsilon_{b}\right)}$
we use the identity, $\frac{d\left(\varepsilon_{a}-\varepsilon_{b}\right)}{d\varepsilon_{a}}=1+\nu_{\textnormal{FeSe}}$ to transform between strain derivatives. Here, $\nu_{\textnormal{FeSe}}$ is the Poisson ratio of FeSe calculated from ultrasound data \cite{Zvyagina2013}. Clearly,  $\Delta\rho(\varepsilon_{2})$ and $2m_{66}\,\rho_t\,(1+\nu_{\textnormal{FeSe}})=\frac{d\Delta\rho}{d\varepsilon_a}$ behave similarly for $T>T_{s}$, explaining experimentally observed tail of $\Delta\rho$ above $T_{s}$. The scaling yields $\varepsilon_{2}=7.7\times10^{-4}$ ($\sim$ 40\% of the distortion in the orthorhombic phase) for the external strain applied through the horseshoe device. Below $T_{s}$, samples in the elastoresistivity setup are not fully detwinned, so that the domains dominate the measured $m_{66}$, which prohibits such a comparison.

To determine the effect of strain on the resistivity below $T_{s}$,
we return to the two resistivity curves at constant strain $\rho_{a}(\varepsilon_{1})$
and $\rho_{a}(\varepsilon_{2})$ in Fig.~2(b) obtained using the horseshoe
device, which fully detwins the samples. In the linear response regime, we can
approximate $\frac{d\Delta\rho}{d\varepsilon_{a}}\approx\frac{\Delta\rho(\varepsilon_{2})-\Delta\rho(\varepsilon_{1})}{\varepsilon_{2}-\varepsilon_{1}}$. The derivative is used to extract the intrinsic resistivity anisotropy between the $a$ and $b$ directions of a single-domain sample in the absence of external strain, $\Delta\rho(\varepsilon=0)\approx\Delta\rho(\varepsilon_{2})-\frac{\Delta\rho(\varepsilon_{2})-\Delta\rho(\varepsilon_{1})}{\varepsilon_{2}-\varepsilon_{1}}\,{\varepsilon_{2}}$.
The constant value of $\frac{\varepsilon_{2}}{\varepsilon_{2}-\varepsilon_{1}}\approx 3.7$ is fixed by enforcing $\Delta\rho(\varepsilon=0)=0$
in the tetragonal state. The resulting $\Delta\rho(\varepsilon=0)$ (red line in Fig.~2(b)) clearly displays a broad maximum 20 K below $T_{s}$.

\begin{figure}
\includegraphics[width=8.6cm]{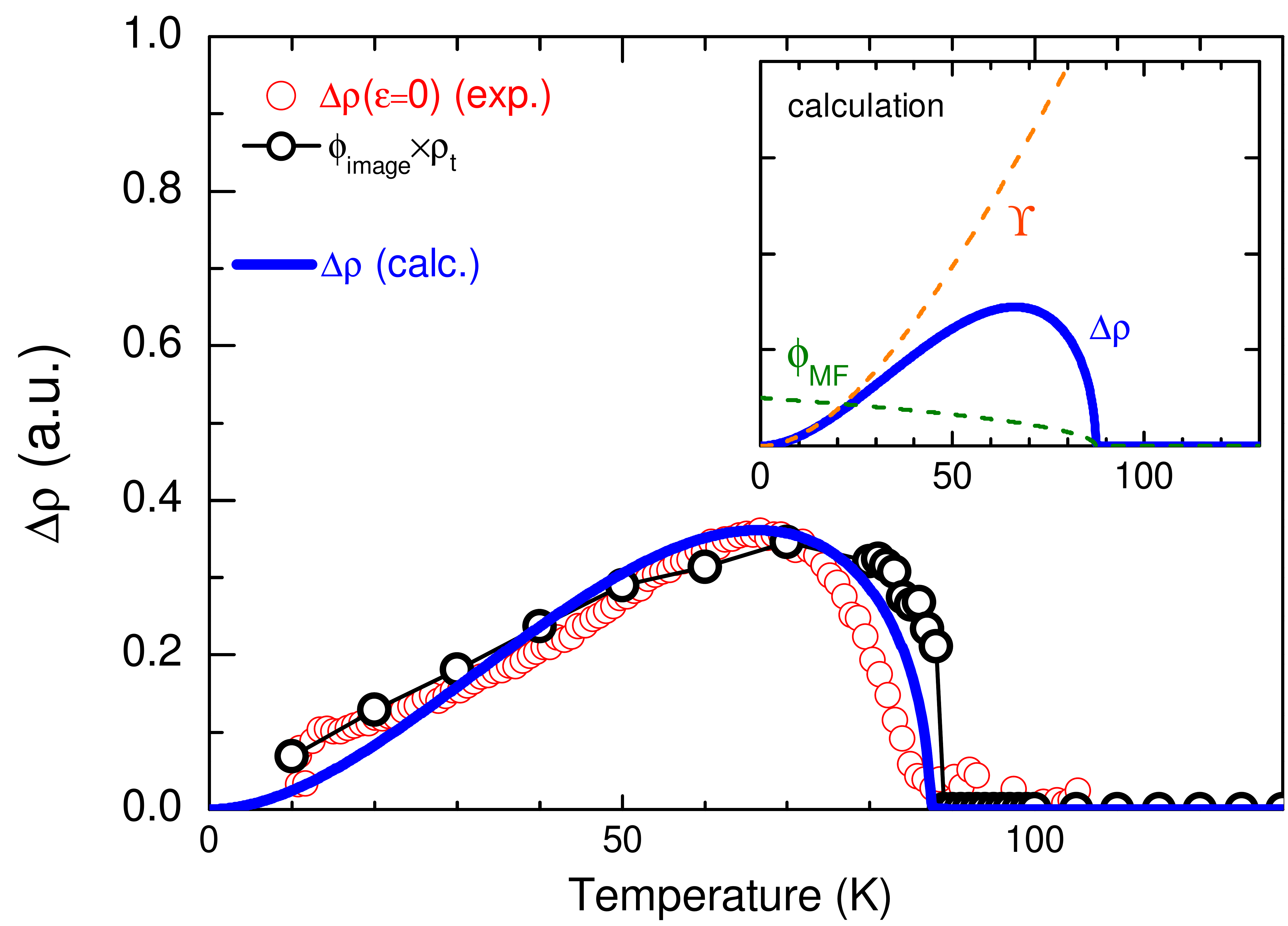} \protect\protect\caption{(color online) Experimental temperature-dependent resistivity anisotropy in the
zero-strain limit (from Fig. 2 (b), red symbols), $\Delta\rho(T)$, compared to the
product of the temperature-dependent experimentally determined order
parameter $\phi_{\mathrm{image}}$ of Fig. \ref{imaging}(g) and the
isotropic resistivity $\rho_{t}$ of Fig. \ref{resistivity}(a) (black circles). The
blue line shows the results of a model calculation where $\Delta\rho$
is determined by the product of a mean-field-type order nematic parameter
$\phi\left(T\right)$ and a temperature dependent function $\Upsilon(T)$
resulting from inelastic scattering promoted by spin fluctuations
(see inset).}

\label{summary}
\end{figure}

The previous comparison between $m_{66}$ and $\Delta \rho$ reveals that, above $T_s$, the resistivity anisotropy is proportional to the strain, and therefore to the nematic order parameter $\phi$. A similar behavior was experimentally observed in the iron pnictides \cite{stress}. This relationship indeed is more general: because $\Delta \rho$ and $\phi$ break the same symmetry, they are generally proportional to each other, i.e. $\Delta \rho = \Upsilon \phi$, were $\Upsilon$ is the proportionality factor \cite{Chu2012,Schutt15}.
It is clear from Fig.~\ref{imaging}
that $\phi$ displays a standard order-parameter behavior, monotonically
increasing upon cooling. In contrast, the resistivity
anisotropy, $\Delta\rho$, shows a pronounced peak below $T_{s}$ and decreases to
 nearly zero at $T\rightarrow0$. This behavior must
therefore arise from the temperature-dependence of the proportionality
factor, $\Upsilon$, which should also vanish as $T\rightarrow0$,
since $\phi$ remains finite and large at $T\rightarrow0$. Among the possible
microscopic contributions to $\Delta\rho$ affecting $\Upsilon$ --  anisotropies of Fermi surface, elastic and inelastic
scattering rates, only the
latter  one naturally leads to the observed behavior. Phenomenologically,
this is nicely illustrated by using $\phi_{\mathrm{image}}$ of Fig.~\ref{imaging}
as a proxy of $\phi$, and the resistivity of the twinned sample,
$\rho_{t}(T)$, as a proxy of $\Upsilon$. The latter relies on the
assumption that the inelastic scattering also dominates the isotropic
transport properties. Indeed, the product $\phi_{\mathrm{image}}(T)\rho_{t}(T)$,
shown by the black symbols in Fig.~\ref{summary}, captures much
of the temperature dependence of $\Delta\rho(T)$.

In order to develop a microscopic scenario for this behavior, we consider
the three-band model of Ref.~\cite{Rafaelcalc}, in which electrons
are scattered by magnetic fluctuations. This model contains one circular
hole pocket at the center of the Brillouin zone and two electron pockets
centered at momenta $\left(\pi,0\right)$ and $\left(0,\pi\right)$.
Below $T_{s}$, the onset of nematic order leads to stronger fluctuations
at the ordering vector $\left(\pi,0\right)$ than at $\left(0,\pi\right)$,
a behavior observed experimentally by the neutron scattering \cite{Dai14}. Depending on the relative positions of the
hot spots -- points of the hole and electron pockets connected by
the ordering vectors -- one finds $\rho_{a}>\rho_{b}$ or $\rho_{b}>\rho_{a}$
(see also Ref.~\cite{Timm14}). Indeed, the change in the positions
of the hot spots from hole-doping to electron-doping was argued
in Ref. \cite{ErickNature} as a possible reason for the sign change
of $\Delta\rho$. Using the formalism developed
in Ref.~\cite{Rafaelcalc}, we perform an expansion of the resistivity
anisotropy, finding $\Delta\rho=\Upsilon\phi$. Here, we assume $\phi\left(T\right)$
to display a mean-field like behavior $\phi\left(T\right)=\phi_{0}\sqrt{1-T/T_s}$.
The proportionality constant $\Upsilon$, arising from the scattering of electrons by magnetic excitations, depends on the magnetic correlation
length, $\xi$, and on the Landau damping of the magnetic fluctuations,
$\Gamma$. In particular, we find $\Upsilon=\Upsilon_{0}T\left(1+3\Gamma\xi^{-2}/2\pi T\right)^{-1}$,
where $\Upsilon_{0}$ is a constant that depends on the geometry of
the Fermi surface and on the residual resistivity. This leads to $\Upsilon(T\rightarrow0)\sim T^{2}$,
and therefore, the different temperature dependencies of $\phi(T)$ and
$\Upsilon(T)$ gives rise to a maximum in $\Delta\rho$ below $T_{s}$.
This behavior is illustrated in Fig.~\ref{summary}, where we plot
the calculated $\Delta\rho$ for $\xi=3a$ and $\Gamma=150$ meV (the product $\phi_{0} \Upsilon_{0}$ is treated as a fitting
parameter to the data). Note that $\xi$ was assumed to be small
and temperature-independent above $T_{s}$, in agreement with the
NMR data \cite{Boehmer2015,Baek2015}. Below $T_{s}$, the
onset of nematic order renormalizes $\xi$ and leads to its enhancement
\cite{Zhang15}, as observed in the same NMR data. The good agreement
between the calculated and the measured $\Delta\rho$ suggests that the inelastic scattering by anisotropic magnetic fluctuations can explain
the experimentally observed non-monotonic temperature-dependence of
the in-plane resistivity anisotropy.

In conclusion, the comparison of direct transport and elastoresistivity
measurements
in FeSe was used to extract the
intrinsic in-plane resistivity anisotropy of strain-free samples. Strong non-monotonic temperature dependence, displaying a maximum
below $T_{s}$ and becoming very small as $T\to0$ limit was observed. This behavior
is explained by anisotropic inelastic scattering as a main contribution to $\Delta\rho$, shedding new light
on the origin of nematicity in iron-based superconductors.

\begin{acknowledgments}
The experimental work was supported by the U.S. Department of Energy
(DOE), Office of Basic Energy Sciences, Division of Materials Sciences
and Engineering. The experimental research was performed at Ames Laboratory,
which is operated for the U.S. DOE by Iowa State University under
Contract No.~DE-AC02-07CH11358. M. S. acknowledges the support from
the Humboldt Foundation. R. M. F. is supported by the U.S. Department
of Energy, Office of Science, Basic Energy Sciences, under Award No.
DE-SC0012336.
G. D. was funded by the Gordon and Betty Moore Foundation's EPiQS Initiative through Grant GBMF4411.
\end{acknowledgments}


\begin{thebibliography}{10}

\bibitem{detwinning} M. A. Tanatar, E. C. Blomberg, A. Kreyssig,
M. G. Kim, N. Ni, A. Thaler, S. L. Bud'ko, P. C. Canfield, A. I. Goldman,
I. I. Mazin and R. Prozorov, \prb~ \textbf{81}, 184508
(2010).

\bibitem{Fisher1} J-H. Chu, J. G. Analytis, K. De Greve, P. L. McMahon,
Z. Islam, Y. Yamamoto, and I. R. Fisher, Science 329, 824 (2010).

\bibitem{Chen} J. J. Ying, X. F. Wang, T. Wu, Z. J. Xiang, R. H.
Liu, Y. J. Yan, A. F. Wang, M. Zhang, G. J. Ye, P. Cheng, J. P. Hu,
and X. H. Chen, \prl~ \textbf{107}, 067001 (2011).

\bibitem{ErickNature} E. C. Blomberg, M. A. Tanatar, R. M. Fernandes,
I. I. Mazin, Bing Shen, Hai-Hu Wen, M. D. Johannes, J. Schmalian,
and R. Prozorov, Nature Comm. \textbf{4}, 1914 (2013).

\bibitem{Chen2} J. Q. Ma, X. G. Luo, P. Cheng, N. Zhu, D. Y. Liu, F. Chen, J. J. Ying, A. F. Wang, X. F. Lu, B. Lei, and X. H. Chen
Phys. Rev. B {\bf 89}, 174512 (2014)


\bibitem{Miyasaka} Tatsuya Kobayashi, Kiyohisa Tanaka, Shigeki Miyasaka,
Setsuko Tajima, J. Phys. Soc. Jpn. \textbf{84}, 094707 (2015).

\bibitem{Uchida} Taichi Terashima, Nobuyuki Kurita, Megumi Tomita,
Kunihiro Kihou, Chul-Ho Lee, Yasuhide Tomioka, Toshimitsu Ito, Akira
Iyo, Hiroshi Eisaki, Tian Liang, Masamichi Nakajima, Shigeyuki Ishida,
Shin-ichi Uchida, Hisatomo Harima, and Shinya Uji, \prl~
\textbf{107}, 176402 (2011).

\bibitem{Fisher14} Hsueh-Hui Kuo and Ian R. Fisher, \prl~
\textbf{112}, 227001 (2014).

\bibitem{FisherReview} I. R. Fisher, L. Degiorgi, and Z.X. Shen, Rep.
Progr. Phys. \textbf{74}, 124506 (2011).

\bibitem{FernandesNaturereview} R. M. Fernandes, A. V. Chubukov, J. Schmalian,
Nature Phys. \textbf{10}, 97 (2014).

\bibitem{Devereaux10} C.-C. Chen, J. Maciejko, A. P. Sorini, B. Moritz, R. R. P. Singh, and T. P. Devereaux, \prb~ \textbf{82},
100504 (2010).

\bibitem{Phillips11} Weicheng Lv and Philip Phillips, Phys. Rev.
B \textbf{84}, 174512 (2011).

\bibitem{Dagotto12} S. Liang, G. Alvarez, C. Sen, A. Moreo, and E.
Dagotto, \prl~ \textbf{109}, 047001 (2012).

\bibitem{Davis13} M. P. Allan, T-M. Chuang, F. Massee, Yang Xie,
Ni Ni, S. L. Bud'ko, G. S. Boebinger, Q. Wang, D. S. Dessau, P. C.
Canfield, M. S. Golden, and J. C. Davis, Nature Phys. \textbf{9},
220 (2013).

\bibitem{Andersen} Maria N. Gastiasoro, I. Paul, Y. Wang, P. J. Hirschfeld,
Brian M. Andersen, \prl~ \textbf{113}, 127001 (2014).

\bibitem{Rafaelcalc} Rafael M. Fernandes, Elihu Abrahams, and J\"org
Schmalian, \prl~ \textbf{107}, 217002 (2011).

\bibitem{Timm14} Maxim Breitkreiz, Philip M. R. Brydon, and Carsten
Timm, \prb~ \textbf{90}, 121104(R) (2014).

\bibitem{Dai14} Xingye Lu, J. T. Park, Rui Zhang, Huiqian Luo, Andriy
H. Nevidomskyy, Qimiao Si, and Pengcheng Dai, Science \textbf{345},
657 (2014).

\bibitem{Mirri2014} C. Mirri, A. Dusza, S. Bastelberger, J.-H. Chu,
H.-H. Kuo, I. R. Fisher, and L. Degiorgi, \prb~ \textbf{90},
155125 (2014).

\bibitem{Mirri2015} C. Mirri, A. Dusza, S. Bastelberger, M. Chinotti,
L. Degiorgi, J.-H. Chu, H.-H. Kuo, and I. R. Fisher, \prl~
\textbf{115}, 107001 (2015).

\bibitem{Valenzuela10} B. Valenzuela, E. Bascones, and M. J. Calderon,
\prl~ \textbf{105}, 207202 (2010).


\bibitem{Dirac} Khuong K. Huynh, Yoichi Tanabe, and Katsumi Tanigaki,
\prl~ \textbf{106}, 217004 (2011).

\bibitem{Fisherdiracnematicity} Hsueh-Hui Kuo, Jiun-Haw Chu, Scott
C. Riggs, Leo Yu, Peter L. McMahon, Kristiaan De Greve, Yoshihisa
Yamamoto, James G. Analytis, and Ian R. Fisher, \prb~ \textbf{84},
054540 (2011).



\bibitem{FeSe} Fong-Chi Hsu, Jiu-Yong Luo, Kuo-Wei Yeh, Ta-Kun Chen,
Tzu-Wen Huang, Phillip M. Wu, Yong-Chi Lee, Yi-Lin Huang, Yan-Yi Chu,
Der-Chung Yan, Maw-Kuen Wu, Proc Natl. Acad. Sci. U S A. \textbf{105},
14262 (2008).


\bibitem{rrrFeSe} Shigeru Kasahara, Tatsuya Watashige, Tetsuo Hanaguri,
Yuhki Kohsaka, Takuya Yamashita, Yusuke Shimoyama, Yuta Mizukami,
Ryota Endo, Hiroaki Ikeda, Kazushi Aoyama, Taichi Terashima, Shinya
Uji, Thomas Wolf, Hilbert von L\"{o}hneysen, Takasada Shibauchi, and Yuji
Matsuda, Proc. Nat. Acad.Sci. \textbf{111} 16309 (2014).

\bibitem{McQueen2009} T. M. McQueen, A. J. Williams, P. W. Stephens,
J. Tao, Y. Zhu, V. Ksenofontov, F. Casper, C. Felser, and R. J. Cava,
\prl~ \textbf{103}, 057002 (2009).


\bibitem{Boehmer2013} A. E. B\"ohmer, F. Hardy, F. Eilers, D. Ernst,
P. Adelmann, P. Schweiss, T. Wolf, C. Meingast, \prb~ \textbf{87},
180505(R) (2013).


\bibitem{imaging} M. A. Tanatar, A. Kreyssig, S. Nandi, N. Ni, S.
L. Bud'ko, P. C. Canfield, A. I. Goldman, and R. Prozorov, \prb~
\textbf{79}, 180508 (R) (2009).


\bibitem{stress} E. C. Blomberg, A. Kreyssig, M. A. Tanatar, R. M.
Fernandes, M. G. Kim, A. Thaler, J. Schmalian, S. L. Bud'ko, P. C.
Canfield, A. I. Goldman, and R. Prozorov, \prb~ \textbf{85},
144509 (2012).

\bibitem{FeTe} Juan Jiang, C. He, Y. Zhang, M. Xu , Q. Q. Ge, Z.
R. Ye, F. Chen, B. P. Xie, and D. L. Feng, \prb~ \textbf{88},
115130 (2013).

\bibitem{detwinningPdoped} H.-H. Kuo, James G. Analytis, J.-H. Chu, R. M. Fernandes, J. Schmalian, and I. R. Fisher, Phys. Rev. B {\bf 86}, 134507 (2012).

\bibitem{Chu2012} J-H. Chu, H.-H. Kuo, J. G. Analytis, I. R. Fisher, Science 337, 710 (2012).

\bibitem{nematicsusceptibility} Hsueh-Hui Kuo, Maxwell C. Shapiro,
Scott C. Riggs, and Ian R. Fisher, \prb~ \textbf{88}, 085113
(2013).

\bibitem{Watson2015} M. D. Watson, T. K. Kim, A. A. Haghighirad,
N. R. Davies, A. McCollam, A. Narayanan, S. F. Blake, Y. L. Chen,
S. Ghannadzadeh, A. J. Schofield, M. Hoesch, C. Meingast, T. Wolf,
and A. I. Coldea, \prb~ \textbf{91}, 155106 (2015).



\bibitem{Zvyagina2013} G.~A.~Zvyagina, T.~N.~Gaydamak, K.~R.~Zhekov,
I.~V.~ Bilich, V.~D.~ Fil, D.~A.~ Chareev, and A.~N.~Vasiliev,
Europhys. Lett. \textbf{101}, 56005 (2013).

\bibitem{Schutt15} Michael Sch\"utt and Rafael M. Fernandes, \prl~ \textbf{115}, 027005 (2015).

\bibitem{Boehmer2015} A. E. B\"ohmer, T. Arai, F. Hardy, T. Hattori,
T. Iye, T. Wolf, H. v. L\"ohneysen, K. Ishida, and C. Meingast, \prl~ \textbf{114}, 027001 (2015).

\bibitem{Baek2015} S.~H.~Baek, D.~V.~Efremov, J.~M.~ Ok, J.~S.~
Kim, Jeroen van den Brink, and B.~ B\"uchner, Nature Materials \textbf{14},
210 (2015).

\bibitem{Zhang15} Qiang Zhang, Rafael M. Fernandes, Jagat Lamsal,
Jiaqiang Yan, Songxue Chi, Gregory S. Tucker, Daniel K. Pratt, Jeffrey
W. Lynn, R. W. McCallum, Paul C. Canfield, Thomas A. Lograsso, Alan
I. Goldman, David Vaknin, and Robert J. McQueeney, \prl~
\textbf{114}, 057001 (2015).


\end{thebibliography}
\end{document}